**Polariton waves in nonlinear dielectric medium**

I V Dzedolik, O S Karakchieva

Taurida National V I Vernadsky University, 4, Vernadsky Avenue, 95007, Simferopol, Ukraine
E-mail: dzedolik@crimea.edu

**Abstract**

The phonon-polariton spectrum in dielectric medium with the third order nonlinearity was theoretically obtained. Dependence of number of polariton spectrum branches on intensity of electromagnetic field was investigated. The appearance of new branches located in the polariton spectrum gap was caused by the influence of dispersion of the third order dielectric susceptibility at increment of the field intensity in the medium. The soliton and cnoidal wave solutions for the polariton excitations for these new spectrum branches were obtained. The all-optical logic gates OR and NOT are proposed as an example of the theory application.

## 1. Introduction

The properties of polaritons that are the quanta of mixed states of electromagnetic waves and dipole excitations of medium (phonons, excitons, plasmons) are still actively investigated in different dielectric, magnetic and semiconductor mediums [1 - 20]. Nonlinear effects appear at generation of polaritons when the frequency of electromagnetic wave is close to resonance frequency lying in terahertz range at a crystal lattice and to electron resonant frequency lying near infrared or optical ranges [3 - 10, 12]. It leads to enhancement of the vibration amplitude, i.e. to the nonlinear oscillations of ions and electrons. Nonlinear effects also become apparent when the nonlinear electron response of medium arises as a result of a strong electromagnetic wave scattering [1 - 4, 6 - 10, 12, 14, 18]. The bulk and surface phonon-polaritons in the dielectric medium, plasmon-polaritons attract attention of researchers for design of controllable filters, all-optical logic gates, signal delay lines and other devices of the microwave and optical circuits [6, 9, 11, 12, 16, ].

The nonlinear dynamics of optical pulse in spectral region close to resonance frequency of longitudinal exciton was explored in the paper [7]. The behavior of low and upper branches of polariton spectrum in nongyrotropic cubic Kerr-type crystal CuCl was investigated. It was shown that polaritons of the upper branch could propagate as solitons forming the "potential". This potential could trap the polaritons of low branch and it led to their propagation through the crystal at the same frequency. In the paper [8] the influence of the polariton dynamics at Kerr-like nonlinearity on dynamics of optical pulses in the dielectric medium close to polariton gap was investigated. The authors showed that inside the spectrum gap there were no solutions in the form of plane waves, but the solutions in the form of solitary waves took place both inside and outside the polariton gap. However, the authors of the paper [8] did not consider the dispersion of the third order susceptibility of medium. Polariton spectra were received in multilayered nonlinear medium in the paper [10], where the authors showed that the number of branches of the spectrum changed dependently on the wavevector, phonon frequency, thickness and nonlinearity of each layer. The new technique of measuring terahertz reflection response by phonon-polariton wave was developed in the paper [13]. The phonon-polariton propagation in nonaxial one-dimensional piezoelectrical superlattice was studied in the paper [15]. Magnetophonon-polaritons in the superlattice were investigated in the paper [17]. In the papers [14, 19 - 20] the properties of phonon-polariton waves, masses of polaritons, dependence of polariton spectrum on intensity of the electromagnetic field and intensity of the external electric field were investigated in nonlinear dielectric mediums and waveguides. The experimental observations of interference between atomic spin coherence and optical field generating polariton waves in controllable time-delayed beamsplitter with dynamically tunable splitting ratio were described in the paper [22].

The above mentioned papers are devoted to analysis of polariton properties and it indicates the common interest of researches in this field of physics.

In this paper we theoretically investigate the properties of phonon-polariton spectrum in nonlinear dielectric medium with the third order susceptibility , i.e. Kerr-type nonlinear medium. In this medium the polaritons represent the bound states of photons and phonons. We show how intensity of electromagnetic field influences the number of polariton spectrum branches. The enhancement of electromagnetic field intensity leads to increasing the nonlinear response of dielectric medium and it results in  appearance of the additional branches in the polariton spectrum gap. New branches of polariton spectrum have harmonical, cnoidal and soliton solutions for polariton waves. To demonstrate this effect we have got the phonon-polariton spectra in nonlinear medium considering the dispersion of the third order nonlinear susceptibility for the first harmonic of wave frequency.

## 2. Theoretical model

Consider the theoretical model of the classical electromagnetic field interaction with the ions forming a crystal lattice. If the electromagnetic wave propagates in the dielectric crystal we can describe this process using the system of equations:

1) equation of the ion motion in a unit cell of the crystal lattice

$$m_{eff}\frac{d^2\mathbf{R}}{dt^2} + m_{eff}\Gamma\frac{d\mathbf{R}}{dt} + \nabla_R U_R = e_{eff}\left(\mathbf{E} + \frac{1}{c}\frac{d\mathbf{R}}{dt}\times\mathbf{B}\right) \qquad (1)$$

where $e_{eff}, m_{eff}$ are the effective ion charge and mass in the lattice cell, $\mathbf{R} = \mathbf{r}_+ - \mathbf{r}_-$ is the displacement vector of positive and negative lattices of ions, $U_R = (q_{1R}/2)R^2 + (q_{2R}/3)R^3 + (q_{3R}/4)R^4$ is the potential energy of ions, $\Gamma$ is the damping factor;

2) equation of  the outer shell electron motion in the ion

$$m\frac{d^2\mathbf{r}}{dt^2} + m\Gamma\frac{d\mathbf{r}}{dt} + \nabla_r U_r = -e\left(\mathbf{E} + \frac{1}{c}\frac{d\mathbf{r}}{dt}\times\mathbf{B}\right) \qquad (2)$$

where $U_r = (q_{1r}/2)r^2 + (q_{2r}/3)r^3 + (q_{3r}/4)r^4$ is the potential energy of electron;

3) the electromagnetic field equations

$$\nabla\times\mathbf{B} = c^{-1}(\dot{\mathbf{E}} + 4\pi\dot{\mathbf{P}}), \quad \nabla\times\mathbf{E} = -c^{-1}\dot{\mathbf{B}}, \qquad (3)$$

where $\mathbf{P} = e_{eff}N_C\mathbf{R} - eN_e\mathbf{r}$ is the polarization vector of medium, $N_C$ is the number of cells in the unit of volume, $N_e$ is the number of electrons in the unit of volume, $q_{jr}, q_{jR}$ are the phenomenological elastic parameters of the medium, the overdot denotes partial time derivative. In the system of equations (1)-(3) we take into account the bound of charges by electromagnetic field.

We can neglect the response of the magnetic component of the high-frequency electromagnetic field $|\mathbf{E}| >> |c^{-1}(d\mathbf{R}/dt)\times\mathbf{B}|$, $|\mathbf{E}| >> |c^{-1}(d\mathbf{r}/dt)\times\mathbf{B}|$ in the medium. Then we represent solutions of the motion equations (1) and (2) as the series where index of term is the order of infinitesimal

$\mathbf{r} = \mathbf{r}_0 + \mathbf{r}_1 + \mathbf{r}_2 + \mathbf{r}_3$, $\mathbf{R} = \mathbf{R}_0 + \mathbf{R}_1 + \mathbf{R}_2 + \mathbf{R}_3$. If the electromagnetic field is harmonic $E \sim e^{-i\omega t}$, it is easy to obtain the polarization vector of medium by the method of sequential approximations [4]. We can obtain the polarization vector by this method in the form of

$$\mathbf{P} = \chi_1 \mathbf{E}_a e^{-i\omega t} + \chi_{20} E_a \mathbf{E}_a + \chi_{22} E_a \mathbf{E}_a e^{-i2\omega t} + \chi_{31} E_a^2 \mathbf{E}_a e^{-i\omega t} + \chi_{33} E_a^2 \mathbf{E}_a e^{-i3\omega t}, \qquad (4)$$

where $\chi_1 = \frac{1}{4\pi}\left(\frac{\omega_e^2}{\tilde{\omega}_1^2} + \frac{\omega_I^2}{\tilde{\Omega}_1^2}\right)$, $\chi_{20} = \frac{1}{4\pi}\left(\frac{e\alpha_{2r}\omega_e^2}{m\omega_0^2\left((\omega_0^2-\omega^2)^2+\omega^2\Gamma^2\right)} - \frac{e_{eff}\alpha_{2R}\omega_I^2}{m_{eff}\Omega_\perp^2\left((\Omega_\perp^2-\omega^2)^2+\omega^2\Gamma^2\right)}\right)$,

$\chi_{22} = \frac{1}{4\pi}\left(\frac{e\alpha_{2r}\omega_e^2}{m(\tilde{\omega}_1^2)^2\tilde{\omega}_2^2} - \frac{e_{eff}\alpha_{2R}\omega_I^2}{m_{eff}(\tilde{\Omega}_1^2)^2\tilde{\Omega}_2^2}\right)$, $\chi_{31} = -\frac{1}{4\pi}\left(\frac{e^2\alpha_{3r}\omega_e^2}{m^2(\tilde{\omega}_1^2)^3(\tilde{\omega}_1^2)^*} + \frac{e_{eff}^2\alpha_{3R}\omega_I^2}{m_{eff}^2(\tilde{\Omega}_1^2)^3(\tilde{\Omega}_1^2)^*}\right)$

$\chi_{33} = -\frac{1}{4\pi}\left(\frac{e^2\alpha_{3r}\omega_e^2}{m^2(\tilde{\omega}_1^2)^3\tilde{\omega}_3^2} + \frac{e_{eff}^2\alpha_{3R}\omega_I^2}{m_{eff}^2(\tilde{\Omega}_1^2)^3\tilde{\Omega}_3^2}\right)$ are the linear and nonlinear susceptibilities of medium,

$\tilde{\omega}_1^2 = \omega_0^2 - \omega^2 - i\Gamma\omega$, $\tilde{\omega}_2^2 = \omega_0^2 - (2\omega)^2 - i2\Gamma\omega$, $\tilde{\omega}_3^2 = \omega_0^2 - (3\omega)^2 - i3\Gamma\omega$, $\tilde{\Omega}_1^2 = \Omega_\perp^2 - \omega^2 - i\Gamma\omega$,

$\tilde{\Omega}_2^2 = \Omega_\perp^2 - (2\omega)^2 - i2\Gamma\omega$, $\tilde{\Omega}_3^2 = \Omega_\perp^2 - (3\omega)^2 - i3\Gamma\omega$; $\omega_e^2 = 4\pi e^2 N_e m^{-1}$, $\omega_I^2 = 4\pi e_{eff}^2 N_C m_{eff}^{-1}$ are the electron and ion plasma frequencies; $\omega_0^2 = q_{1r} m^{-1}$ is the electron resonance frequency, $\Omega_\perp^2 = q_{1R} m_{eff}^{-1}$ is the resonance frequency of lattice; $\alpha_{2r} = q_{2r} m^{-1}$, $\alpha_{3r} = q_{3r} m^{-1}$, $\alpha_{2R} = q_{2R} m_{eff}^{-1}$, $\alpha_{3R} = q_{3R} m_{eff}^{-1}$; $\Gamma, q_{jr}, q_{jR}$ are the phenomenological parameters depending on linear and nonlinear properties of medium. Properties of the given medium also are described by electron $\omega_e$ and ion $\omega_I$ plasma frequencies.

Consider the medium with centers of inversion, i.e. the medium with response of the third order susceptibility $\chi_3$. We represent the electromagnetic field as the set of plane waves and consider the interactions only at the first harmonic $\mathbf{E} = \mathbf{E}_a \exp(-i\omega t + i\mathbf{k}\mathbf{r})$. Then the polarization vector of the medium has the form $\mathbf{P} = (\chi_1 + \chi_{31}E_a^2)\mathbf{E}$. In this case the permittivity of medium is defined by the expression

$$\varepsilon = 1 + 4\pi\chi_1 + 4\pi\chi_{31}E_a^2, \qquad (5)$$

In the expression for permittivity (5) both the electron and ion responses of the electromagnetic field medium are considered.

Then we eliminate the magnetic inductance vector **B** from the field equations (3) $\nabla \times \nabla \times \mathbf{E} = -c^{-2}\varepsilon \ddot{\mathbf{E}}$, and get the system of algebraic equations $(k^2 - c^{-2}\varepsilon\omega^2)\mathbf{E} = \mathbf{k}(\mathbf{kE})$. We resolve the electric field vector into transverse and longitudinal components $\mathbf{E} = \mathbf{E}_\perp + \mathbf{E}_{//}$ relating to the wavevector **k**. Assumed that the interaction of electromagnetic waves and charges in the medium effects by transverse field component $\mathbf{E}_\perp$, that is $\mathbf{kE}_\perp = 0$, we obtain the dispersion equation for polaritons in nonlinear medium

$$k^2 - c^{-2}\omega^2\left(1 + 4\pi\chi_1 + 4\pi\chi_{31}E_a^2\right) = 0. \tag{6}$$

## 3. Polariton spectrum

The potential energy of ions and electrons can be approximated in the crystal of cubic system by nonlinear functions

$$U_R = exp(-R)\left[q(R^3+R)^{-1} - R^{-1}\right] - const \approx (q_{1R}/2)R^2 + (q_{2R}/3)R^3 - (q_{3R}/4)R^4 - const,$$

$$U_r = exp(-r)\left[q(r^3+r)^{-1} - r^{-1}\right] - const \approx (q_{1r}/2)r^2 + (q_{2r}/3)r^3 - (q_{3r}/4)r^4 - const.$$

The susceptibility of the third order is more than zero $\chi_{31} > 0, \chi_{33} > 0$, because $\alpha_{3R} = -q_{3R}m_{eff}^{-1} < 0$, $\alpha_{3r} = -q_{3r}m^{-1} < 0$ are negative in this case (see the expression (4)).

The polariton spectrum depends on the density of electromagnetic field $\sim E_a^2$ in the Kerr-type nonlinear medium. In the weakly nonlinear medium at $4\pi\chi_{31}E_a^2 \approx 0$ the spectrum of polaritons has only three branches (fig. 1a), and this result agrees with the deduction in [3]. In the nonlinear medium at $4\pi\chi_{31}E_a^2 = 10^{-5}$ the polariton spectrum has nine branches (fig. 1b). In this case the spectrum still has branches 1, 2, 3, but six new branches with numbers 4, 5, 6 and 7, 8, 9 appear. New branches 4 and 5 coincide with themselves and smoothly go up, but the branch 6 has a weak declination down. The branches 7, 8, 9 have the same behavior: the branch 7 has the weak declination down, and branches 8, 9 smoothly go up and completely coincide. Polariton spectrum in weakly nonlinear medium has only two gaps, but as the electromagnetic field intensity in nonlinear medium increases, the third gap appears: the first gap (between branch 2 and branches 4, 5, 6), the second gap (between branch 2 and branches 7, 8, 9) and the third gap (between a branch 3 and branches 7, 8, 9).

The appearance of new branches in the polariton spectrum is caused by the dispersion of the third order dielectric susceptibility of medium $\chi_{31}(\omega)$ at increasing of the electromagnetic field intensity $\sim E_a^2$ (see the equation (4)). In weakly nonlinear medium the dispersion equation (6) has the sixth degree of the frequency $\omega$, in more nonlinear medium the dispersion equation (6) has the eighteenth degree of the frequency because the susceptibility of the third order possesses the dispersion.

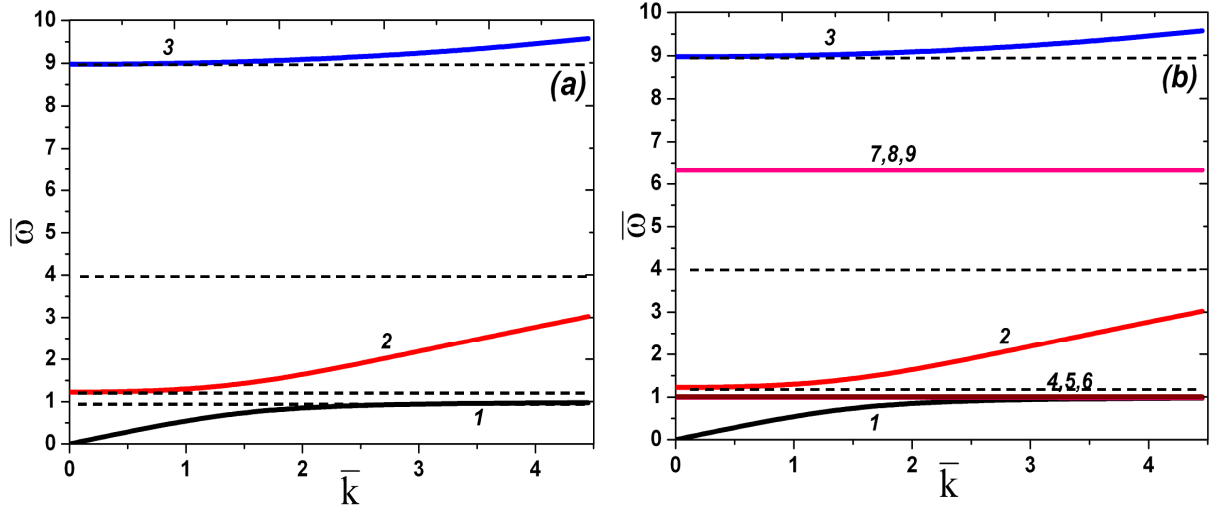

Fig. 1 (Color online): The polariton spectrums in nonlinear medium with dispersion of the third order susceptibility $\chi_3$ – (a) at $4\pi\chi_{31}E_a^2 \approx 0$; (b) $4\pi\chi_{31}E_a^2 = 10^{-5}$. In the Fig. 1 $\Gamma = 0$; $\bar{\omega} = \omega/\Omega_\perp, \bar{k} = ck/\Omega_\perp$ are the dimensionless values. The solid lines are the branches of polariton spectrum; the dashed lines are the limits of spectrum gaps.

The curves have been obtained numerically by solving the dispersion equation (6) for polariton spectrum. For example, we can use the medium and wave parameters $\Omega_\perp \sim 10^{13} s^{-1}$, $k = 3 \cdot 10^2 ... 3 \cdot 10^3 cm^{-1}$ to calculation of spectra in the Fig. 1. If the wavevector increases $k \to 10^5 cm^{-1}$, the third spectrum gap is widen.

**4. Spectrum gap polaritons**

It is known that a plane harmonic wave is unstable in nonlinear medium [9, 23]. Instability of the wave leads to its modulation as nonlinear periodic (cnoidal) or solitary waves at the corresponding relations of field and medium parameters. Let us view a process of forming of the cnoidal wave or spatial soliton from the harmonic wave with the frequency $\omega_7 \equiv \omega$ that lies at the polariton spectrum gap in nonlinear infinite dielectric medium with the third order of susceptibility (fig. 1b).

We can write the equation of polariton wave as

$$\nabla \times \nabla \times \mathbf{E} + c^{-2}\ddot{\mathbf{E}} = -c^{-2}4\pi \ddot{\mathbf{P}}, \qquad (7)$$

where $\mathbf{P} = \chi_1 \mathbf{E} e^{-i\omega t} + \chi_{31}|E|^2 \mathbf{E} e^{-i\omega t}$. Let us assume that the electric field in infinite dielectric medium looks like $\mathbf{E} = \mathbf{1}_x E_x(x,y,z) + \mathbf{1}_y E_y(x,y,z)$, then from (7) we obtain the equation system

$$\frac{\partial^2 E_x}{\partial y^2} + \frac{\partial^2 E_x}{\partial z^2} - \frac{\partial^2 E_y}{\partial x \partial y} + \frac{\omega^2}{c^2}(1+4\pi\chi_1)E_x + \frac{4\pi\omega^2 \chi_{31}}{c^2}\left(|E_x|^2 + |E_y|^2\right)E_x = 0,$$

$$\frac{\partial^2 E_y}{\partial x^2} + \frac{\partial^2 E_y}{\partial z^2} - \frac{\partial^2 E_x}{\partial x \partial y} + \frac{\omega^2}{c^2}(1+4\pi\chi_1)E_y + \frac{4\pi\omega^2 \chi_{31}}{c^2}\left(|E_x|^2 + |E_y|^2\right)E_y = 0, \qquad (8)$$

where expressions for $\chi_1$ and $\chi_{31}$ are given after expression (4). We present the particular solutions of equation system (8) as the plane waves $E_{x,y} = E_{x0,y0} \exp(iq_x x + iq_y y + ikz)$. Then we equate the determinant of system of the algebraic equations for $E_{x0}, E_{y0}$ to zero $(c^{-2}\omega^2\varepsilon - q_x^2 - k^2)(c^{-2}\omega^2\varepsilon - q_y^2 - k^2) - q_x^2 q_y^2 = 0$ and obtain the dispersion equation

$$k^4 - \left(2\frac{\omega^2}{c^2}\varepsilon - k_\perp^2\right)k^2 + \frac{\omega^2}{c^2}\varepsilon\left(\frac{\omega^2}{c^2}\varepsilon - k_\perp^2\right) = 0, \qquad (9)$$

where $\varepsilon = 1 + 4\pi\chi_1 + 4\pi\chi_{31}w$ is the nonlinear permittivity, $w = E_{x0}^2 + E_{y0}^2$ is the electromagnetic energy density, $k_\perp^2 = q_x^2 + q_y^2$. We can evaluate the values of wavevector by solving the dispersion equation (9)

$$k_+ = \frac{\omega}{c}(1 + 4\pi\chi_1 + 4\pi\chi_{31}w)^{1/2},$$

$$k_- = \left[\frac{\omega^2}{c^2}(1 + 4\pi\chi_1 + 4\pi\chi_{31}w) - k_\perp^2\right]^{1/2}. \qquad (10)$$

The linear $\chi_1$ and nonlinear $\chi_{31}$ susceptibilities of medium are complex (see expression (4)), and the wavevectors (10) are complex too, that's why the modulation instability of the plane wave in this case takes place. But in the transparent medium with $\Gamma = 0$ the susceptibilities $\chi_1, \chi_{31}$ will have real values, and instability of the plane wave will depend on a sign before $\chi_{31}$ and electromagnetic energy densities $w$ [9].

We present the polariton wave propagating along the axis $z$ as the carrier harmonic. Then modulation of this wave can be described by the field as $E_{x,y} = \tilde{E}_{x,y}(x,y,z)\exp(ikz)$, where $\tilde{E}_{x,y}(x,y,z)$ is the slowly varying amplitude along the axis $z$. We can neglect the second derivative on $z$ from amplitude and from set (8) we obtain the system of nonlinear Schrodinger equations

$$i2k\frac{\partial \tilde{E}_x}{\partial z} + \frac{\partial^2 \tilde{E}_x}{\partial y^2} - \frac{\partial^2 \tilde{E}_y}{\partial x \partial y} + \alpha_3\left(\left|\tilde{E}_x\right|^2 + \left|\tilde{E}_y\right|^2\right)\tilde{E}_x = 0,$$

$$i2k\frac{\partial \tilde{E}_y}{\partial z} + \frac{\partial^2 \tilde{E}_y}{\partial x^2} - \frac{\partial^2 \tilde{E}_x}{\partial x \partial y} + \alpha_3\left(\left|\tilde{E}_x\right|^2 + \left|\tilde{E}_y\right|^2\right)\tilde{E}_y = 0,$$

(11)

where $\alpha_3 = 4\pi c^{-2}\omega^2 \chi_{31}$ ($\alpha_3 > 0$ at $\chi_{31} > 0$ in the self-focusing medium, and $\alpha_3 < 0$ at $\chi_{31} < 0$ in the self-defocusing medium). The system of equations (11) describes the nonlinear periodic and solitary polariton waves in nonlinear medium with the third order susceptibility. In the special case at the absence of mixed derivative of cross members, the system of equations (11) describes the vector spatial solitons in infinite dielectric nonlinear medium [9].

Let us assume that the vector polariton wave does not change the form propagating along the axis $z$. We define the dependence of field on longitudinal coordinate by the constant phase displacement $q$, $\tilde{E}_j = e_j \exp(iqz)$, where $e_j$ is the real amplitude, $j = x, y$. In this case we obtain the system of equations from (11)

$$\frac{\partial^2 e_x}{\partial y^2} - \frac{\partial^2 e_y}{\partial x \partial y} + \alpha_1 e_x + \alpha_3\left(e_x^2 + e_y^2\right)e_x = 0,$$

$$\frac{\partial^2 e_y}{\partial x^2} - \frac{\partial^2 e_x}{\partial x \partial y} + \alpha_1 e_y + \alpha_3\left(e_x^2 + e_y^2\right)e_y = 0,$$

(12)

where $\alpha_1 = c^{-2}\omega^2(1 + 4\pi\chi_1) - k^2 - 2kq$. We can get the solutions of equation system (12) with the boundary conditions $e_j \to 0$, $de_j/dr_j \to 0$ at $|r_j| \to \infty$. Substituting into equation system (12) the partial solutions in the form of plane inhomogeneous waves $e_x = e_{0x}\exp(-k_x|x| - k_y|y|)$, $e_y = e_{0y}\exp(-k_x|x| - k_y|y|)$ we obtain the dispersion equation for the case of weak nonlinearity $\alpha_1 \gg \alpha_3(e_x^2 + e_y^2)$,

$$c^{-2}\omega^2(1 + 4\pi\chi_1) - k^2 - 2kq + k_x^2 + k_y^2 + \alpha_3(e_x^2 + e_y^2) = 0.$$

(13)

The views of envelopes $e_x$ and $e_y$ of vector polariton wave are represented in fig. 2.

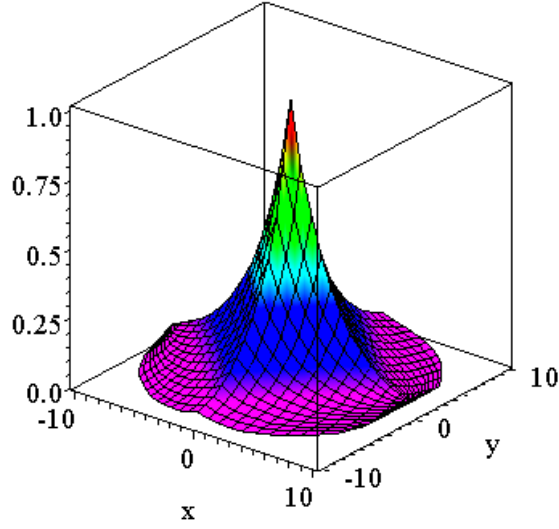

Fig. 2. The envelopes $e_x$ or $e_y$ of inhomogeneous vector polariton wave.

The system of equations (12) is split on two uncoupled equations at linear polarization of the wave (along the axis $x$ or axis $y$)

$$\frac{d^2 e_x}{dy^2} - \alpha_1 e_x + \alpha_3 e_x^3 = 0,$$
$$\frac{d^2 e_y}{dx^2} - \alpha_1 e_y + \alpha_3 e_y^3 = 0,$$
(14)

where $\alpha_1 = 2kq$ if the dispersion law $c^{-2}\omega^2(1+4\pi\chi_1) - k^2 = 0$ takes place. The sign "plus" before the nonlinear term characterizes the self-focusing medium, where the light solitons can form [9].

We obtain the solution of an equation (14) for light polariton soliton. The first integral of the equation looks like $(de/d\xi)^2 = \alpha_1 e^2 - \alpha_3 e^4/2 + C$, where $C$ is a integration constant. The boundary conditions for the soliton $e \to 0$, $de/d\xi \to 0$ at $|\xi| \to \infty$ allow to define the integration constant $C = 0$, and boundary conditions for soliton centre $e(0) = const$, $de(0)/d\xi = 0$, disposed in the point $\xi = 0$, allow to define the phase displacement $q = \alpha_3 e^2(0)/4k$. The second integral of the equations for light soliton looks like $\sqrt{|\alpha_3|/2}\,\xi = \int e^{-1}(e^2(0) - e^2)^{-1/2} de$, and after integration we obtain $e(\xi) = e(0)\,sech\!\left(e(0)\sqrt{|\alpha_3|/2}\,\xi\right)$. In this case the polariton wave looks like the space soliton with stationary form (fig. 3a) polarized on the axis $x$ or $y$,

$$\tilde{E}_x(y) = e_x(0) \cosh^{-1}\left(\frac{e_x(0)\sqrt{|\alpha_3|}}{\sqrt{2}} y\right) \exp\left(i\frac{\alpha_3 e_j^2(0)}{4k} z\right), \tag{15}$$

$$\tilde{E}_y(x) = e_y(0) \cosh^{-1}\left(\frac{e_y(0)\sqrt{|\alpha_3|}}{\sqrt{2}} x\right) \exp\left(i\frac{\alpha_3 e_j^2(0)}{4k} z\right). \tag{16}$$

Besides the spatial solitons there are the cnoidal waves in the self-focusing medium with $\alpha_3 > 0$. The cnoidal wave can transform to the soliton in the special case [14]. We choose the boundary conditions $e = const$, $de/d\xi = 0$ at $|\xi| \to \infty$ for cnoidal wave, i.e. the integration constant will not be equal zero $C = \alpha_3 e_\infty^4/2 - \alpha_1 e_\infty^2 = const$. Then the second integral looks like $\sqrt{|\alpha_3|/2}\,\xi = \int_0^e (C' + \alpha' e^2 - e^4)^{-1/2} de$, where $\alpha' = 2\alpha_1 \alpha_3^{-1}$, $C' = 2C\alpha_3^{-1}$, and phase displacement is equal $q = \alpha_3 e_\infty^2/4k - C/2k e_\infty^2$. In the self-focusing medium we obtain the envelope of cnoidal polariton wave with polarization along the axis $x$ or axis $y$ as the elliptic cosine,

$$\tilde{E}_x(y) = \tilde{e}_x\, cn\left(K_x y, \tilde{k}_x\right) \exp(iq_1 z), \tag{17}$$

$$\tilde{E}_y(x) = \tilde{e}_y\, cn\left(K_y x, \tilde{k}_y\right) \exp(iq_2 z), \tag{18}$$

where $\tilde{e}_j = \left[\alpha'/2 + (\alpha'^2/4 + C'_j)^{1/2}\right]^{1/2}$, $a_j = (\alpha'^2/4 + C'_j)^{1/4}\sqrt{|\alpha_3|}$, $q_{1,2} = \alpha_3 e_{\infty(x,y)}^2/4k - C/2k e_{\infty(x,y)}^2$, $\tilde{k}_j = \left[2 + \alpha'(\alpha'^2/4 + C'_j)^{-1/2}\right]^{1/2}/2$ is the modulus of elliptic integral. The view of envelope of cnoidal polariton waves in the self-focusing medium is represented in fig. 3b. At $\tilde{k} \to 1$ the elliptic cosine is transformed to the hyperbolic secant $cn(\xi,1) \to cosh^{-1}(\xi)$ describing the spatial soliton, i.e. in the case if $C \to 0$.

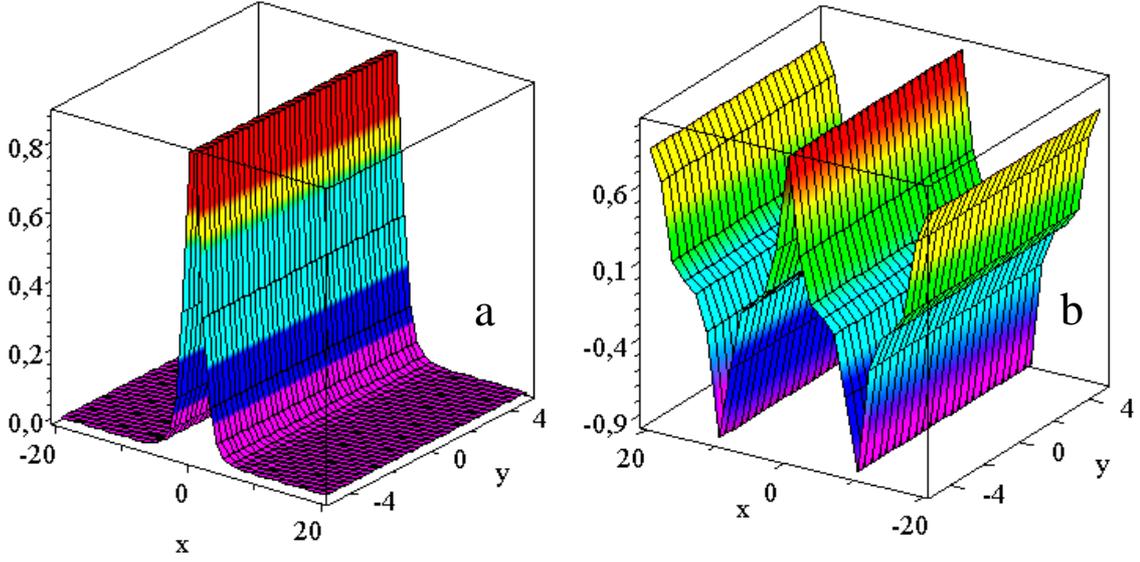

Fig. 3. The envelope of linearly polarized polariton wave $e_y(x)$ in the self-focusing medium:

a) as the spatial soliton, b) as the cnoidal wave.

In the self-defocusing medium at $\alpha_3 < 0$ we obtain the envelope of polariton waves as the elliptic tangent

$$e(\xi) = \tilde{e}\, tn(K\xi, \tilde{k}_{-1}), \qquad (19)$$

where $\tilde{e} = \left[\alpha'/2 - (\alpha'^2/4 - C')^{1/2}\right]^{1/2}$, $K = \left[-\alpha'/2 + (\alpha'^2/4 - C')^{1/2}\right]^{1/2} \sqrt{|\alpha_3|/2}$,

$\tilde{k}_{-1} = \left[2(\alpha'^2/4 - C')^{1/2} - \alpha'\right]^{1/2} \left[-\alpha'/2 + (\alpha'^2/4 - C')^{1/2}\right]^{-1/2}$. The solution (19) as the cnoidal wave is unstable in the self-defocusing medium; it can be explained by properties of the elliptic tangent.

## 5. Application of examined model

The examined model serves to explain the processing of nonlinear optical filter with sharp frequency transparency. In the frequency gap $\Delta\overline{\omega} = 9 - 4 = 5$ only a wave harmonic with normalized frequency $\overline{\omega} = \omega/\Omega_\perp = 6.4$ of the wave packet passes through the nonlinear medium (Fig. 1) if the packet intensity is $E_a^2 \sim (4\pi\chi_{31} \cdot 10^5)^{-1}$. The remainder harmonics of the wave packet do not pass through this medium. None of the packet harmonic has not pass otherwise if $E_a^2 \ll (4\pi\chi_{31} \cdot 10^5)^{-1}$. The polariton wave in the gap can be excited as the harmonical wave, cnoidal wave or spatial polariton, and after passing through the medium the polariton wave transforms to the radiation in air at terahertz or optical ranges.

Some crystals and glasses with symmetric structure of lattices are the applicable mediums with necessary properties for arising of nonlinear polaritons, for example [9], the silicate glass $SiO_2$ with $\chi_1 = 0.037$, $\chi_{31} = 2.4 \cdot 10^{-16} cm^2/W$; $AlGaAs$ with $\chi_{31} = 2.0 \cdot 10^{-13} cm^2/W$; crystal polydiacetylene paratoluene sulfate with $\chi_{31} = 2.2 \cdot 10^{-12} cm^2/W$ at wavelength $\lambda = 1.6 \mu m$ and another mediums. The units $(esu)$ correlate with units $(SI)$ as $\chi_{31}(esu) = (9/4\pi) \cdot 10^9 \chi_{31}(m^2/W)$. Various experiments with nonlinear electromagnetic wave propagation in the bulk mediums with third order nonlinearity were described in the monograph [9]. Our model of nonlinear polariton waves allows predicting the new effects of wave propagation in bulk media and their applications. Particularly, the polaritons of new branches of the spectrum with frequencies $\omega_7$, $\omega_8$ or $\omega_9$, (Fig. 1), which have appeared in polariton spectrum gap due to polaritons decay with frequency $\omega_3$, can be used as the signals in logic gates.

As an example of the theory applications we consider the logic gates OR and NOT based on the nonlinear medium properties mentioned above (Fig. 4). The input signals are $A$ and $B$, the output signal is $Out$. Suppose that the signal with frequency $\omega_3$ represents the logic unit 1, and the signal with frequency $\omega_7$ represents the logic zero 0 (see Fig. 1).

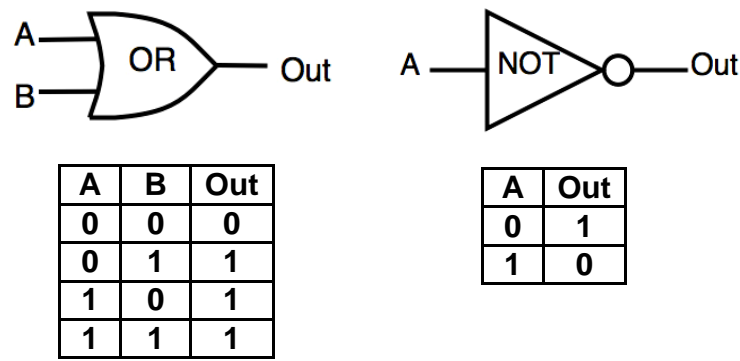

Fig. 4. Logic symbols with truth tables.

The logic gate OR can be designed as the Y-coupler containing three rectangular dielectric waveguides conjugated together (Fig. 5a). The nonlinear polariton pulses representing the signals $A$ or $B$ propagate through the waveguides of Y-coupler without the in-channel interference, because the frequencies of carrier waves of signals logic 0 and logic 1 are different.

The logic gate NOT includes the dispersing prism, rectangular dielectric waveguides, two nonlinear microresonators with mirror faces for signals with frequencies $\omega_3$ and $\omega_7$, and the source of pump waves (Fig. 5b). The signal $A$ representing logic 1 deviates more than signal representing logic 0 in the prism, because the frequency $\omega_3$ is higher than frequency $\omega_7$. Consequently, the signal with frequency $\omega_3$, or the signal with frequency $\omega_7$, passes through its own resonator, where it mixes with two powerful pump waves with total frequency $2\omega_P = \omega_3 + \omega_7$ and wavevector $2k_P = k_3 + k_7$. Four-wave mixing takes place in the resonator $de_3/dt = \gamma_3 e_P^2 e_7$, $de_7/dt = \gamma_7 e_P^2 e_3$, $e_P^2 \approx const$, where $\gamma_{3,7} = const$. After passing through the resonator with frequency $\omega_3$ the signal logic 0 with frequency $\omega_7$ transforms into the signal logic 1 with frequency $\omega_3$, and after propagating through the resonator with frequency $\omega_7$ the signal logic 1 with frequency $\omega_3$ transforms into the signal logic 0 with frequency $\omega_7$; their amplitudes increase as $e_{3,7} \sim exp\left(e_P^2 \sqrt{\gamma_3 \gamma_7} t\right)$.

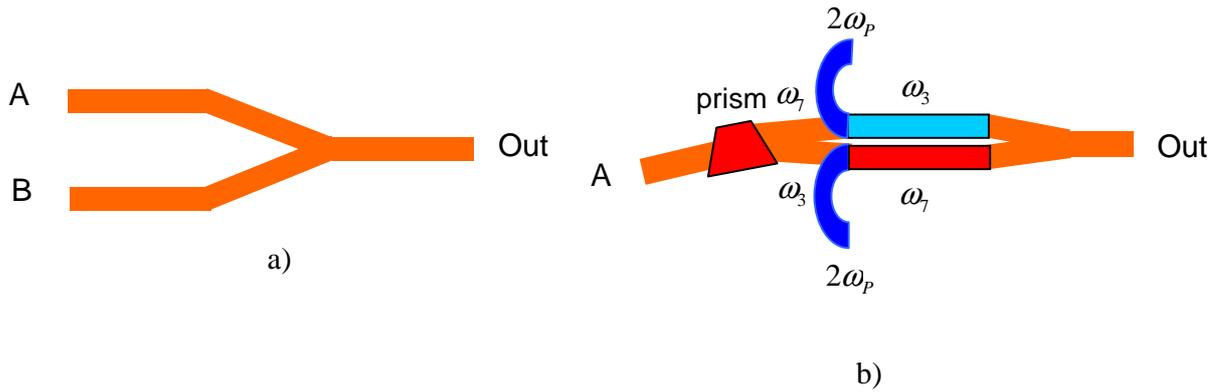

Fig. 5. Logic gates: a) – OR, b) – NOT.

## 6. Conclusion

We have shown theoretically that the dispersion dependence of the third order susceptibility in nonlinear dielectric medium leads to appearance of new branches in polariton spectrum, particularly, in the spectrum gap, as the electromagnetic field intensity increases. These properties of the polariton spectrum in nonlinear dielectric medium may be used for designing controllable optical filters, all-optical logic gates, etc.

It is necessary to note that new branches in the polariton spectrum arise as a consequence of the dispersion of the third order medium susceptibility, and they are represented as the solutions in the form of plane waves. Since the plane wave is unstable in nonlinear medium it can be transformed to the cnoidal wave or spatial soliton depending on the wave and medium parameters.

The polaritons of new spectrum branches can be used as the signals at logic gates. The design of logic gates OR and NOT based on the nonlinear medium properties is proposed in the paper. One can use these logic gates for the representation of the whole tool base for Boolean operations.

**Acknowledgment**

The authors are grateful to Yuri S. Kivshar for discussions about the paper.